# Transaction handling in COM, EJB and .NET

Ditmar Parmeza and Miraldi Fifo

IDT Department, Mälardalen University, Västerås, Sweden

## ABSTRACT

*The technology evolution has shown a very impressive performance in the last years by introducing several technologies that are based on the concept of component. As time passes, new versions of Component-Based technologies are released in order to improve services provided by previous ones. One important issue that regards these technologies is transactional activity. Transactions are important because they consist in sending different small amounts of information collected properly in a single combined unit which makes the process simpler, less expensive and also improves the reliability of the whole system, reducing its chances to go through possible failures. Different Component-Based technologies offer different ways of handling transactions. In this paper, we will review and discuss how transactions are handled in three of them: COM, EJB and .NET. It can be expected that .NET offers more efficient mechanisms due to the fact of being released later than the other two technologies. Nevertheless, COM and EJB are still present in the market and their services are still widely used. Comparing transaction handling in these technologies will be helpful to analyze the advantages and disadvantages of each of them. This comparison and evaluation will be seen in two main perspectives: performance and security.*

## 1 INTRODUCTION

A transaction represents a unit of work that contains several simple operations. Transactions must always fulfill the ACID principles (Atomicity, Consistency, Isolation, Durability) regardless of what technology or platform is used (COM, EJB, .NET etc.). Definitions and explanations of these important four principles are given in the following paragraphs.

**Atomicity** allows transactions to change the state of the system, only if all the participating objects are executed successfully. If any object encounters a problem and it's not executed, then the transaction aborts and none of the changes made to the system are committed. This is known as the aborting process. The aborting and the committing processes need to be done as atomic operations. This characteristic is very useful since the client does not need to recover the system in any case. The whole transaction either succeeds or fails. If it succeeds, then everything is ok. If it fails, then the client can make a new request. [2]

**Consistency** means that all the transactions must leave the system in a consistent state after their execution. This implies that even if the transaction succeeds, it is not guaranteed that there is consistency. The transaction has to ensure that the system is in a consistent state even after its execution. The system should pass from a consistent state (before the transaction) to another consistent state (after the transaction). In case that any error occurs, the client must be able to abort the transaction and roll back the system in the previous consistent state. [2]

**Isolation** means that while the transaction is being executed, no one can see the changes made from it during the intermediate phase. This happens because this intermediate phase is not consistent. This principle is essential for the consistency of the system. All the resources that are used from one transaction cannot be accessed from other transactions until the transaction which is using them, aborts or succeeds. In a few words, the resources lock the data from the other transactions while they are used from another transaction. There can be deadlocks. A deadlock occurs when two transactions each compete for accessing a resource that actually is used from the other transaction. The solution is to abort both the contenders. COM+ uses serialization for transactions. The serialization means that the result of concurrent transactions is the same as if they would be executed serially (one after another). To achieve this, all the resources that are used from a transaction during its execution, are not accessible for the others. [2]

**Durability** means that all the changes made from the transactions to the system should be persistent. This is achieved through the durable storages such as file-systems, magnetic tapes or optical storages. All the changes must be sent to one of these storages, because a crash of the machine can occur at any moment and the memory can be erased. If this happens and all the changes of the transaction were in the memory, then all of them can get lost and the system enters in an inconsistent state. The new consistent state of the system after a successful transaction must be stored in a durable resource which can survive to a crash of the machine. One solution to this problem is to keep log files. These files help the system to recover and make sure that the changes get completed. [2]

The rest of the content of this paper is organized as following:

In section 2, we make an introduction for each technology. A general overview of COM, EJB and .NET is presented in this section but no details about transaction management are given. Section 3 is used to explain transaction handling in COM (COM+). The mechanisms provided by EJB to handle transactions are presented in section 4. In the successive section, we talk about transactions in .NET. In section 6, we compare the mechanisms and ways of handling transactions by the three technologies. The comparison is seen in the performance and security perspectives. Section 7 contains conclusions that are derived from the comparison made in the previous section. Section 8 is about future work i.e., we discuss issues about transaction handling in COM, EJB and .NET that are left to be improved in the future.

## 2 A GENERAL OVERVIEW OF COM, EJB AND .NET

The first technology that we are going to discuss is COM (Component Object Model). This technology uses binary components. Distributed Component Object Model (DCOM) provides COM with distribution features and Microsoft Transaction Server (MTS) provides DCOM with persistency and transactional services. All of them together make what is called COM+. *A COM interface is seen as a C++ virtual class and takes the form of a list of data and function declarations without associated code*. [1]

COM uses the concept of polymorphism in terms of allowing a double indirection to access its interface elements. COM objects can create or find interfaces by using a protocol that is provided from the model. The main interface is *IUnknown* and its basic functions are *QueryInterface, AddRef* and *Release*. A COM object is a piece of binary code. These objects can be packaged in two ways: executable files or DLL (Dynamically Linked Libraries). [1]

COM uses two ways for the problem of composition. The first one is the containment. It implies that an object contains other objects. It implements all the interfaces belonging to the contained objects by calling the same function for each of them. The second way is the aggregation which is more complex. The object container provides the contained object with the interface, without implementing it. In order to be compatible with the *IUnknown* interface, both the container object and the contained object have to adapt their source code. The framework of COM consists of many standard interfaces and a runtime that collaborates with DLL for interpreting the calls for objects creations and objects releases for returning the handles of interfaces. The framework in DCOM is extended. COM and COM+ do not support life-cycle concepts since both of them are execution-time and binary component models. [1]

Now, we will talk about Enterprise Java Beans. This technology is distinguished from the others for one basic reason: it is very simple compared to the other technologies. A bean is a set of ports, if we talk from the external (interface) view of it. There are four types of ports: methods, properties, event sources and event sinks (listeners). We can see a bean in two contexts: at composition time or at run-time. We can see how the model looks like in Figure 1. The "property" port can be used for both contexts of the bean. It parameterizes the bean at composition time and it is also an attribute at run-time. Three operations can be related to the property: getting its value, setting a new value of it and editing its actual value. We can see some other elements beside the four types of ports in Figure 1. The "bounded property" generates an event each time it changes its value. The "vetoable" property can change its value when all the components related to it accept this change. The method property is the same as the concept of methods in Java Programming Language. We can just call the method. This is the only possible operation related to the 'method' port. 'Event sources' generate events. 'Event sinkers' receive events. An event source can be connected or disconnected to an event sink. There are multicast and unicast event sources. The first one can be connected to many event sinkers while the second one can be connected to only one event sinker. [1]

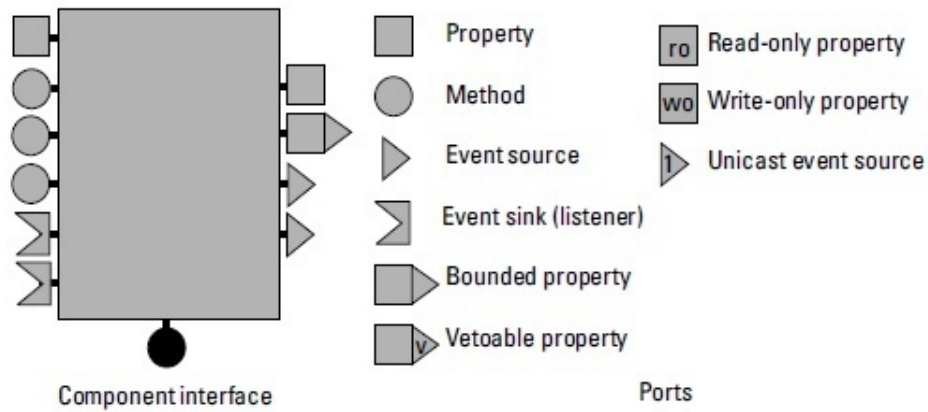

Figure 1.   Interface of a bean component [1]

Bean components are implemented by one Java object. This object is encapsulated inside the bean. Figure 2 shows how the component looks like.

In reality, the complex implementations of the bean components are widely used. Implementation is presented only conceptually in Figure 2. In fact, there are many objects inside the component instead of just one as shown in the figure. The objects inside the components can call other objects outside. One of the most important features of Enterprise JavaBeans (EJB) model is assembly. EJB uses a tool (builder) to assembly the bean components. Figure 3 gives a better idea. It is the assembly tool that manages the number of connections between the components. We can see the combination of Component-Based technologies with traditional technologies in Figure 3 (b).

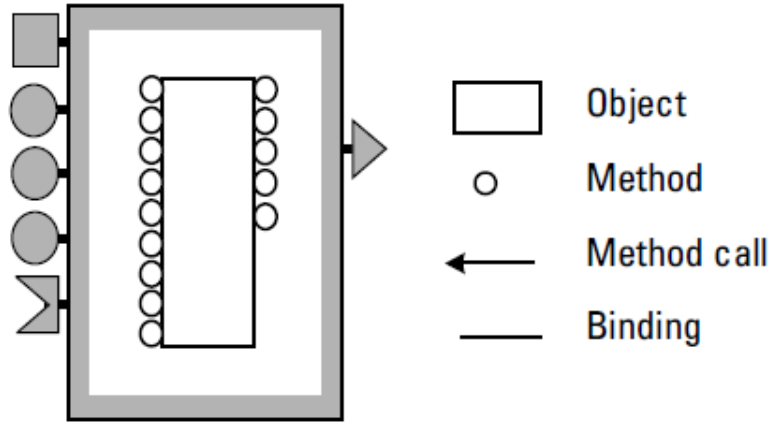

Figure 2. A simple implementation of a bean component [1]

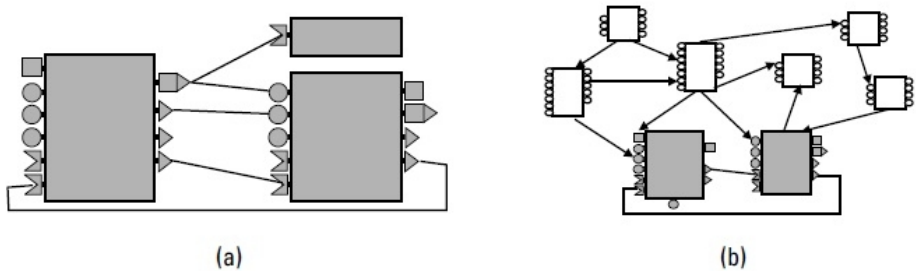

Figure 3. Bean-based systems: (a) components assembly (b) heterogeneous assembly. [1]

As for packaging, EJB uses archives. Therefore, there are several components in one archive. By loading the archive, we load all the components that are packaged inside. In order to avoid problems with file sharing and duplications, EJB uses dependency relationships for the packaged items. Another problem that may occur is the complexity of the component due to its code. In order to avoid this problem, we mark the design phase items so that we can remove them from the final application. [1]

.NET is the last technology that we are going to discuss. This is the newest technology of Microsoft. It does not rely on binary concept of components like COM does. It relies instead on language interoperability and introspection. Because of that, .NET provides us with a language called Microsoft Intermediate Language (MSIL) which has the same functions as the Java Byte Code in interpreting and introspecting. The Common Language Runtime (CLR) has more similarities with a Java Virtual Machine. [1]

In .NET, it is the program itself that contains all the necessary information for the relationships between the components. This is the programming language approach. More specifically, it is the compiler that must provide this kind of information. .NET uses the "manifest'" to gather all the information needed for an assembly. This technology uses C# as a programming language. The compiler generates the code, but it also generates the interface description of the component in the "manifest". This interface is called "assembly" and it is a list of import and export types. There is also a dynamic linker that provides the connections of the resources during the execution time.

A component is constituted of modules that can be executable files or DLLs. The list of these modules is provided in the command line of the compiler during the compilation of the main module. The main module and the "manifest" are in the same file. The modules can be loaded whenever they are required. .NET model is not a hierarchical one since the modules cannot be assembled. Figure 4 gives an idea of how the interface and the implementation of the component in .NET work. [1]

The framework in .NET is the language run-time support. Microsoft Transactional Server (MTS) has the responsibility to manage transactions.

In .NET, the assemblies are local for a specific application. Because of this, many DLLs can run in the same time. Each assembly keeps a history of its own versions as well as versions of the other assemblies it depends on. Then, a dynamic loader chooses the most appropriate version based on the information provided from the assembly and on a set of default rules. These characteristics optimize the packaging and the deployment process. [1]

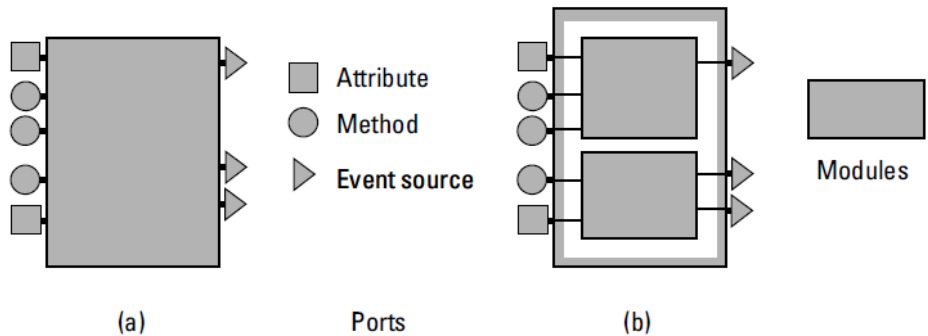

Figure 4.  .NET component (a) interface  (b) implementation  [1]

# 3 TRANSACTION HANDLING IN COM (COM+)

The first attempt or improvement that was made to COM regarding transaction management was MTS (Microsoft Transaction Server). MTS is responsible for handling transactions in COM components. To be more precise, the specialized unit inside MTS for transaction handling is DTC (Distributed Transaction Coordinator). It coordinates transactions that share the same space or the same resource managers. MTS deprives the client of making calls for beginning or ending the transaction. But the clients are not the only ones to be deprived of such a thing. The start and the end of the transactions are also not visible to the MTS components. If it was not for the Microsoft Transaction Server, each component would be obligated to make calls every time the transaction begins or ends (for that component). In these conditions, it would be really difficult to combine different components and put them into the same transaction since each of them would decide independently for the beginning and ending time of its transaction. Fortunately, MTS solves this problem easily and it is its responsibility to combine components into the same transaction in the most convenient way for the system. [8][9]

COM+ (COM extension) uses attributes for transactions in order to determine their type of protection for each object. An object can share the transactions with the caller, can require a new transaction or can even operate without any transaction protection. There are five types of attributes.

**Disabled –** This attribute is used when the component never accesses a resource manager. The object can share the transactions with its caller. When COM components migrate to COM+, then this attribute has to be active in order to ensure the same transactional behavior of the migrated component in its new technology i.e.COM+.

**Not Supported** – The created objects do not participate in a transaction. This means that the object does not interfere with the transaction of the caller and does not begin a transaction by itself. This is the default attribute.

**Supported** – The created objects participate in existing transactions. This attribute can be activated when the client wants the object to share the transactions with its caller but without requiring a new one for itself.

**Required** – The created objects are transactional. The objects participate in the transactions of the caller. If the caller does not have any transaction, then the created object is the new root for later transactions. This attribute is helpful for components that perform resource activities since it offers a transaction protection for these activities.

**Requires new –** The created objects are the roots for the new transactions regardless the status of the transactions of the caller. The new transactions cannot affect the existing transactions of the caller.

As we already mentioned in the ACID principles discussed in the introduction section, there are some resources are used by transactions. When a resource is used from a transaction it falls into the "transaction boundary". All the resources within this boundary share one single transaction. The first object in a transaction boundary is called root. There is only one root in a transaction. All the other objects under the root are called interior objects. In order to be sure that an object is in the right transaction boundary, we map our transactions before we start writing the components. COM+ analyze the transaction's attribute to determine the status of the new object during runtime. The new object can be a root, can be created in an existing transaction or can be created as a non-transactional object. [3]

In order to give the developers more control over the applications, COM+ allows configurable isolation levels of transactions. There are five types of isolation levels: Serialized, Repeatable read, Read Committed, Read Uncommitted, Any. These types of isolation levels differ from each other in the way how they manage data and the resources that are used from the transaction. [4]

There are also some flags that determine the status of the transaction. The *consistent* flag determines if the transaction is either consistent or inconsistent. The *done* flag determines the duration of the transaction. If the transaction does not perform any operation within a certain limit of time, then we have a time-out. After that, it needs to start from the beginning. There is also another interesting feature of COM+. It is called "*Bring your own transaction*" *(BYOT)*. This feature allows the components to inherit external transactions. [5][6]
COM+ also manages automatic transactions. There are of course some conditions that need to be satisfied in order to manage automatic transactions.

Finally, let's see a simple code that defines the attribute of a transaction: [8]

```
public enum TransactionOption
{
    Disabled,
    NotSupported,
    Supported,
    Required,
    RequiresNew
}
```

As we can see, the attributes are included in a constructor that accepts *enum* parameters of *TransactionOption* type. In order to configure our component, we use the code below: [8]

```
[Transaction(TransactionOption.Required)]
public class MyComponent :ServicedComponent
{...}
```

# 4 TRANSACTION HANDLING IN EJB

Enterprise JavaBeans provides more than one solution when we speak about transaction handling. As in other component-based technologies, transaction control in EJB can be either manual or automatic. Both JTA (Java Transaction API) and a unit called EJB container can be responsible for automatic transactions. [10][11][13] JTA helps EJB applications to communicate with several transaction services. In addition, it provides a high-level interface which is compatible with transaction services and controls transaction boundaries (i.e., starting and ending them). [12]

On the other hand, EJB container is the unit that actually contains the entity beans. Therefore, we can state that EJB is responsible for every entity bean and this implies registering them, providing a remote interface for each case and ensure secure transaction coordination. EJB containers are provided by EJB servers. That means that an EJB server can provide one or more containers to coordinate transactions. [12] It is important to regard an EJB entity as a model of business data. In other words, the entity beans in the container are related to the Business Logic layer in the EJB server. They can also have a direct connection with the database. [10] These relations are shown in Figure 5.

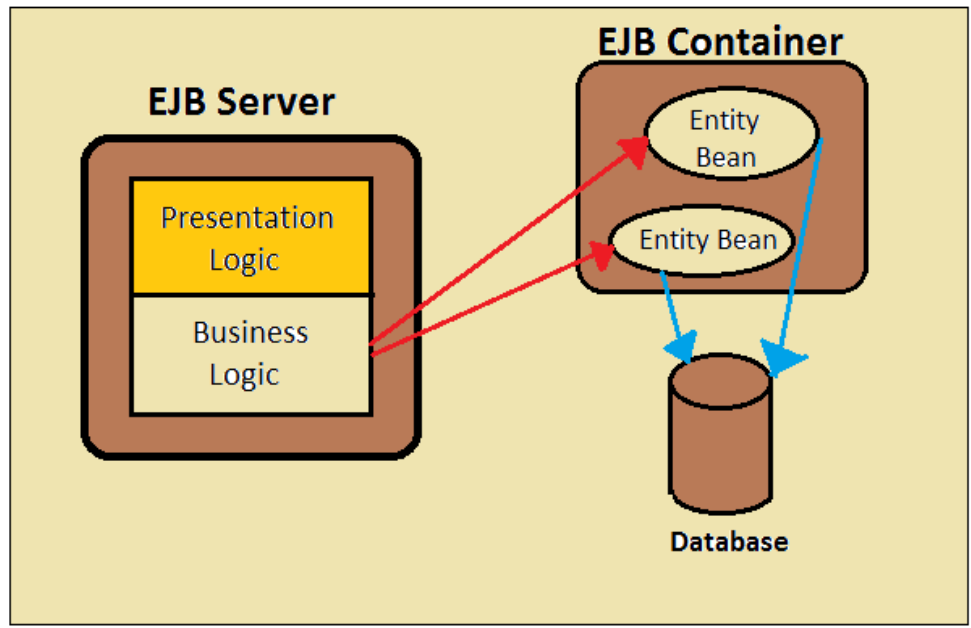

Figure 5. Relations between EJB Server, EJB Container and the Database

Nevertheless, the control of the programmer is very weak meaning that it is not possible for the developer to determine transaction boundaries, commits, rollbacks etc. by written code. They will be automatically determined by the container, as it was stated above. All the responsibility of transaction handling can be also delegated to a database server. [10][12] When we have to deal with simple systems, only one database server is needed and the transactions are local. In the case of more complex systems, more database servers are needed and transactions are distributed or global. [10]

JTS (Java Transaction Service) is another important unit that is very helpful when we have to deal with more complex systems. Basically, this unit makes sure that transactions are coordinated and handled well in several servers. [10][12]

We already admitted that it is the EJB container that controls transactions. Now, we will explain how this control is performed. This is done by declaring some transaction features and attributes in the EJB deployment descriptor.

The declaration of these features and attributes can be done either for the whole EJB or by using methods. [11] The transaction attributes that can be declared, can be set to some values. The possible values are: TX_NOT_SUPPORTED, TX_REQUIRED, TX_SUPPORTS, TX_REQUIRES_NEW, TX_MANDATORY, TX_BEAN_MANAGED. [10][11] We can already see some similarities with transaction attributes in other Component-Based Technologies e.g., COM. The first attribute, TX_NOT_SUPPORTED is used for a bean that cannot perform a transaction. If there is any attempt to do the opposite, the EJB server will throw an exception. TX_REQUIRED means that a bean needs to perform a transaction. If the client has already started a transaction, then that is the exact transaction that will be performed. In the case that no transaction is started from the client, a new transaction will be committed. TX_SUPPORTS is an optional attribute for transactions. Transaction support will be provided only in case that the client has established a transaction context which will be used by the container.

TX_REQUIRES_NEW can be easily confused with TX_REQUIRED but there is a big difference. TX_REQUIRES_NEW implies that a new transaction has to be started by the server and performing this transaction will have a higher priority than performing other ones. That means that even when a transaction has been started by the client, that transaction must be paused and this pause will last as long as it takes for the server to perform its high-priority transaction. After that is completed, the transaction started by the client, can continue. TX_MANDATORY is used when the presence of a transaction started by the client is needed in any case. If the client has not started a transaction, the server will throw the *TransactionRequired* exception. The last attribute, TX_BEAN_MANAGED relies on the code support for transaction handling by a certain EJB. [10][11] Basically, the EJB responsible for controlling the transaction must be coded. A transaction can be controlled by calling and invoking more than just one method in the code. That means that we can begin the transaction by calling one method but other actions i.e. rolling back, committing etc. can be included in other methods. Let's consider the following lines of code: [11]

```
import javax.jts.UserTransaction;
...
EJBContext ic = ...;
...
UserTransaction tx = ic.getUserTransaction();
tx.begin();
... // do work
tx.commit();
```

At first, the interface *javax.jts.UserTransaction* is imported. This way, this interface will be available for the bean thanks to the method *EJBContext. getUserTransaction()*. Basically, every transaction that initiates after TX_BEAN_MANAGED has been invoked, will be suspended by the container. [11]

Another important aspect of transactional activity in EJB is synchronization. [11][14] The idea behind it is that the EJB application gets notifications from the EJB container about the transactional status of the bean. In order to do so, the *javax.ejb.SessionSynchronization* interface must be implemented and imported in the code. There are three types of notifications: *afterBegin()* (notification for the beginning of a new transaction), *beforeCompletion()*(notification that a transaction is about to be committed), *afterCompletion()*(notification that a transaction is done). [11][14]

Finally, we will show a simple transaction illustration in EJB that uses the JTA *javax.transaction.UserTransaction* interface: [11]

```
UserTransaction object = context.getUserTransaction();
try {
        object.begin();
        //Do whatever transaction functionality is necessary
        object.commit();
} catch (Exception ex) {
        try {
                object.rollback();
        } catch (SystemException syex) {
                Throw new EJBException("Rollback failed: " + syex.getMessage());
        }
        throw new EJBException("Transaction failed: " + ex.getMessage());
}
```

An object is initialized by the session bean in order to initiate a transaction context through the *javax.transaction.UserTransaction* interface. Then, it goes through all the main phases by using the right methods i.e., *begin(), commit()* and *rollback()*. Exceptions that are already mentioned in the paragraph regarding transaction attributes, are thrown in each case.

# 5 TRANSACTION HANDLING IN .NET

The fact of being released by Microsoft brings to natural differences between .NET and non-Microsoft technologies e.g., EJB. Another important fact that cannot be ignored is that .NET is relatively considered as a new Microsoft Component-Based technology. Despite of the changes it provided as a latest technology, .NET still remains strongly connected to its predecessors. In this case, predecessors include even the oldest versions of Microsoft Component-Based technologies i.e., COM and especially its extension, COM+. In fact, the relations between .NET technology and COM+ principles and services will be explained in this section. [7][15][17] Either manual or automatic control can be performed in .NET transactions, just like in other technologies. Manual control is enabled by OLE DB, ODBC (old version of OLE DB) or ADO.NET (later version) which have the responsibility to access data. On the other hand, automatic transactions means that the control is performed though a value set in the ASP.NET page or .NET framework class. [7][15] We already mentioned the fact that .NET makes use of COM+ services. Transactional activities are also part of these services. In fact, those components that use these services are considered as serviced components while those that do not are considered as standard managed components. A simple view of a .NET serviced component is shown below: [7]

```
namespace MyNamespace

{
  using System.EnterpriseServices;
  using System.Windows.Forms;//for the MessageBox class

  public interface IMessage
  {
    void ShowMessage(  );
  }
  /// <summary>
  ///   Plain vanilla .NET serviced component
  /// </summary>
  public class MyComponent:ServicedComponent,IMessage
  {
    public MyComponent(  ) {}//constructor
    public void ShowMessage(  )
    {
      MessageBox.Show("Hello!","MyComponent");
    }
  }
}
```

It is important to use the *System.EnterpriseServices* namespace since everything will be inherited from it. In this simple case, the .NET serviced component just implements the *IMessage* interface and displays the message "Hello" by calling the *ShowMessage(  )* method. Inheriting from the class *ServicedComponent* gives a transaction in .NET the same attributes of a transaction in COM+. These transaction attributes include the following options: Disabled, NotSupported, Supported, Required and RequiresNew. These attributes determine if a request for a certain transaction will be fulfilled or rejected. [7][15]

Now, let's take a look at figure 6. Implementation layer, Service layer and the Database layer are shown here. It is the case of performing a distributed transaction between two objects in .NET. [7] A distributed transaction in .NET is conceptually the same as a distributed transaction in other Component-based technologies e.g., EJB. [10] Therefore, more than one database servers are needed in this case in order to perform the distributed transaction and two databases are actually present in the Database layer shown in figure 6. The unit responsible for transaction handling is situated in the Service layer. In this layer, the context for each .NET object must be used but what is most important is the DTC unit. DTC (Distributed Transaction Coordinator) uses mostly the same techniques of handling transactions as other DTCs in other Component-based technologies. That is understandable also because of one major reason. It is not just the DTC part but the whole Service layer that uses services of COM+. This fact is the biggest proof that .NET is so much dependent on COM+. It can be also seen in the figure that the responsibility of transaction handling can be delegated directly to the database server without using the Service layer. The red arrows in the figure represent this connection. In a few words, transactions can be handled either by the DTC in the Service layer or directly by the Database layer.

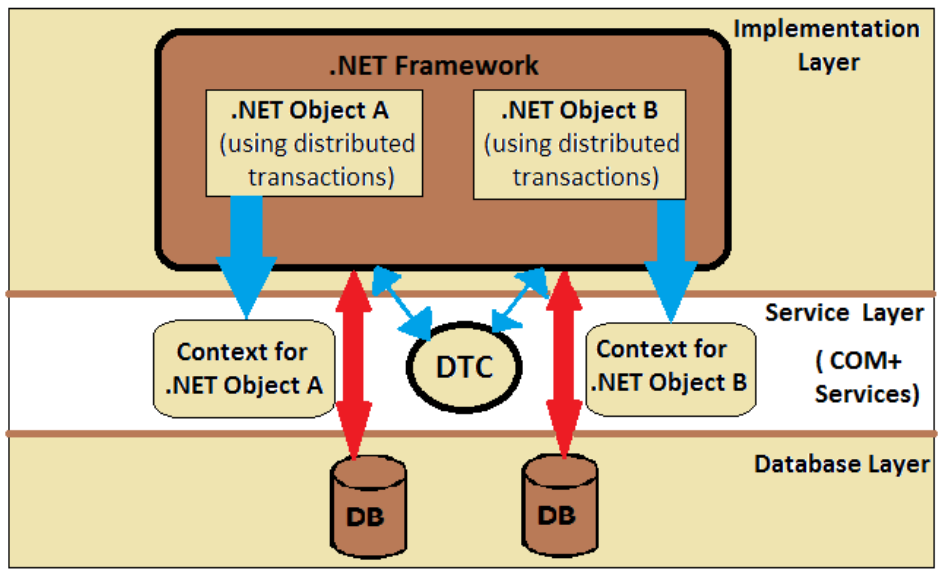

Figure 6. Distributed Transactions in .NET

The earliest versions of .NET framework i.e., .NET Framework 1.0 and .NET Framework 1.1 do not offer a good way to handle distributed transactions. Better solutions are provided by .NET Framework 2.0 [15][16][17] One of the biggest changes is the introduction of two new transaction managers: LTM (Lightweight Transaction Manager) and OLETx Transaction Manager. [16][17] OLETx Transaction Manager works as a MSDTC (Microsoft Distributed Transaction Coordinator) which uses RPC (Remote Procedure Call) for cross-machine calls in order to achieve a higher performance regarding distributed transactions. [17] A good performance is guaranteed also by LTM, the other transaction manager. Another important concept introduced in .NET Framework 2.0 is management namespace. Transactions are now handled though the *System.Transaction* namespace. This is a more efficient way to handle transactions and that is the main reason why it is more preferable than the inheritance from *ServicedComponent*. [7][15][16][17]

Namespace management offers also a higher level of flexibility while handling transactions. That means that it is suitable to handle either manual (explicit) or automatic (implicit) transactions. This is convenient to use even in case of a complicated situation where many resources are needed in order to complete the transactions (distributed transactions). This situation is presented in figure 7 [15].

Resources that are managed using the *System.Transaction* namespace are called *System.Transaction* Resource Managers. Resource managers can be automatically enlisted in a transaction by coding in the client part and detecting the current *TransactionScope* option. *TransactionScope* can be one of the following options: *Required, Requires New* or *Suppress*. *Required* means that the transaction started by the server has a higher priority. *Requires New* represents the beginning of a new transaction. On the other hand, *Suppress* implies not allowing any transaction performing for a certain amount of time.

It is also true that a transaction can be promoted in an automatic way from one technology to another (e.g., from LTM to OleTx) and this is thanks to the *System.Transaction* namespace.

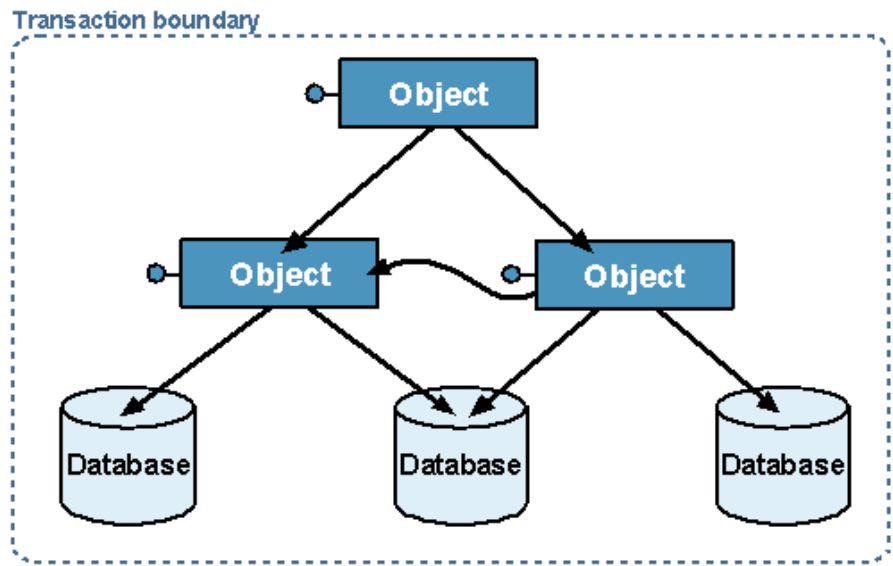

Figure 7. Several resources needed for distributed transactions [15]

Transactions are started by default by a *LightweightTransaction* object. It is up to this object to decide if it will be promoted to a distributed transaction which means that it will be accessed by several resources. [15][17]

*System.Transactions* can be implemented through three different programming models: Declarative, Explicit and Implicit model. In the Declarative model, services inherited from .NET Enterprise are used. At the beginning of each transaction, LTM is chosen by default as transaction manager. It is not a bad model for simple situations i.e., when transactions need to be performed locally and there is only one resource. Despite of this fact, this model presents many difficulties in case of dealing with more complex situations. The Explicit (manual) model requires of course more efforts in terms of code support. Transaction boundaries, commits, rollbacks etc. are specified manually in the code and this way only one transaction can be used many times through several function calls. As for the Implicit (automatic) model, it does not require much effort in coding since all the specifications here are made automatically. Because of this characteristic, the Implicit model remains the most preferable solution. [16][17]

# 6 COMPARISON OF TRANSACTION HANDLING IN COM, EJB AND .NET

In this section, we compare mechanisms and techniques used to manage transactions in COM, EJB and .NET. We focus on two important perspectives regarding transactional activity: performance and security.

## 6.1 Performance

Two experimental studies about the performance of transactional activities in COM, EJB and .NET will be used in this paper. Performance in this context means the amount of transactions that are performed during a certain amount of time. It is logical to say that performance is higher in that case when more transactions can be performed during the same amount of time.

There is one difficulty that this comparison of performance has to face. The experimental studies for COM and .NET are made in the same machine. That means that hardware features and parameters which affect the performance of COM and .NET, are the same. We use reference [18] for the information regarding COM and .NET part. The problem is the EJB part. The data that we need about the EJB transactional performance will be taken from another experimental study [13]. Hardware environment in this case is not exactly the same as the one used for the experimental study of performance in COM and .NET [18]. We have to be conscious that having different hardware environments can make it difficult to us to gather the whole necessary data for the comparison. Nevertheless, a considerable number of experiments are made for the COM and .NET's part. [18] We will try to take into consideration the experiment [18] which is more similar with the one made for EJB [13] in terms of hardware. This way, we will have an idea about the performance in all of the three technologies.

If we analyze the experiment made in [13], we can say that it is applied on the JBoss application, using EJB 2.0 local and the execution time is estimated for 11070 lines of code. We can see from the graph in figure 8, that the time dedicated to the transaction manager starts at 470 ms (approximately) and ends at 920 ms. Therefore, we have 920-470=450 ms as the average time for a transaction. Meanwhile, we refer at pages 56 and 63 in [18], in order to get data for COM and .NET performance. The experiments are made on an SQL server. In this case, we have 10000 lines of code, which is a bit smaller than the number of lines in the EJB case. The experimental studies in [18] made for 100 or 1000 lines of code are not be used in our comparison. We can see that the approximate estimated time for COM+ is 350 ms. On the other hand, it is 310 ms for .NET 2.0 and 270 ms for .NET 3.0. We can put all these values in the graph shown in figure 8. It can be clearly seen from the graph in figure 8 that the execution average time values are close to each other. The value for EJB is a bit higher but that is understandable since it is calculated for 11070 lines of code (1070 more lines than in COM or .NET case).

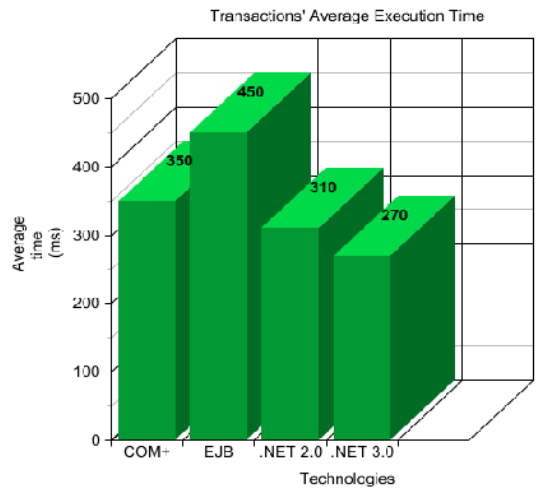

Figure 8. Average execution time of transactions in different technologies

Anyway, we already expected that the performance of COM+ and that of .NET would not differ too much since .NET makes use of COM+ services in a very significant way. It was still expected that .NET would have a slightly higher performance since being a newer technology gives it obvious advantages. A higher performance in .NET 2.0 (40 ms less than COM+) is dedicated to the use of *System.Transaction* namespace which was discussed in section 5. An even better performance is of course achieved by the newest version, .NET Framework 3.0.

## 6.2 Security

Security represents another important issue that we cannot ignore when speaking about transactional activities. In the previous section, we admitted that a better performance can be achieved by the technology that has been released as the latest i.e., .NET. This logic does not work also for security. We cannot state that .NET offers a higher level of security than other Component-Based technologies e.g., COM only because of chronological reasons. Microsoft technologies i.e., COM (COM+) and .NET use role-based security. [7] This type of security makes sure that the client has an appropriate access on a transaction, depending on the role that was given to him. He might have been given a role that allows him to have too much control on the transaction or he might have been given a role that provides him a restricted access to the transaction. In .NET case, the clients are given free choice for the roles. That means that if they confuse the roles, there will be big problems with the security. On the other hand, this does not happen in the case of COM+ role-based security because the roles are determined explicitly by the business domain of the application. Therefore, there is no way to confuse the given roles. [7]

As for EJB, we stated in section 4 that the EJB container is responsible for transaction handling. It can be added that the container has also the responsibility to ensure security in the system. [19] The concept of roles is also used in EJB. These roles are kept in an access list. The container makes sure that every method call is checked automatically and is compared with the role in the access list. If they match, then everything is fine in terms of transactional security. [12]

# 7 CONCLUSIONS

In this paper, we discussed transaction handling in three Component-Based technologies: COM, EJB and .NET. What they have in common is the concept of DTC (Distributed Transaction Coordinator) which handles transactions that are distributed in more than one resource or servers. Two major ways of managing transactions are present in each technology. These ways are known as automatic (implicit) and manual (explicit) management. In the manual management, the developer can determine in a manual way the transaction boundaries and other features that have to do with transactions. That requires significant efforts in coding by the developer. On the other side, everything can be also automated thanks to the transaction attributes. The possible values for transaction attributes are almost the same for COM, EJB and .NET. The only obvious difference is TX_BEAN_MANAGED attribute in EJB which is not present in the two Microsoft technologies.

COM handles transactions thanks to its extensions, MTS and COM+. MTS (Microsoft Transaction Server) include also the DTC which is the actual unit that manages transactions. In EJB, transactions are managed by the EJB container. Apart from the container, JTA (Java Transaction API) is also used to make an automatic management. It provides also a high level interface that is adaptable for other services that regard transactions. In case of dealing with simple systems, the container can be ignored and the responsibility of handling transactions can be passed directly to the database. This happens also in .NET. JTS (Java Transaction Service) is used instead for more complex systems.

As for .NET, it is the latest technology out of the three ones that are taken into consideration in this paper. It uses a considerable amount of COM+ services and this fact makes it depending a lot on COM+. Despite of this fact, .NET has many advantages comparing to the other two technologies. .NET 1.0 and 1.1 offer the possibility of inheritance from *ServicedComponent*. Inheriting helps in facilitating the transactional activity and making it quicker, resulting this way in a higher performance (comparing to COM or EJB). Further improvements are brought by .NET 2.0 which uses *System.Transaction* namespace, making the process of handling transactions more flexible and improving the performance even more.

In order to ensure a proper level of security during a transactional activity, the three technologies use the concept of role. Different roles are given to different clients to determine different levels of access and control over transactions. Despite of being a newer technology, .NET cannot assure a higher security than COM+ because .NET does not offer a solution in case of confusion of roles. On the other hand, EJB uses an access list which keeps track of all the right roles. These roles are later compared with the method calls that are checked automatically. In case that they do not match, an error is generated. This represents a very good security check.

# 8 FUTURE WORK

What is left for future improvements is to use the three technologies that are discussed in this paper in a more sophisticated way. We already admitted that .NET and COM (COM+ services) are strongly connected to each other. This is understandable since they are both Microsoft technologies. Being on the same side, there was no rivalry between them. Therefore, there was no problem on using them together in order to achieve better transactional services.

The real issue is making this kind of collaboration happen between Microsoft (COM, .NET) and non-Microsoft technologies (EJB). The advantages of one technology (e.g., EJB) can be used this way to reduce or eliminate the disadvantages of another one (e.g., COM, .NET). If the transaction management systems in COM and .NET would support the use of EJB components instead of COM components, there would be several positive consequences. For example, EJB requires less effort in coding when we speak about transaction implementation. EJB is also platform independent and COM components do not have this quality. Combining COM, EJB and .NET in an appropriate way will bring to satisfying results in the future.

## Authors

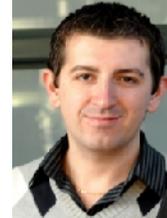

**Ditmar Parmeza** was born on May 26th, 1990 in Shkodër, Albania. He has graduated the Faculty of Information Technology at the Polytechnic University of Tirana in October 2011. He holds a Bachelor diploma in Computer Engineering from 2011 and he completed the first year of the 2-year master program studies in Computer Engineering (also at the Polytechnic University of Tirana). He is currently studying for a 2-year master in Software Engineering at Mälardalen University in Sweden and he has already started working on his Industrial Master Thesis at the E&E (Electronic&Electrical) Architecture Department at Volvo CE (Volvo Construction Equipment) in Eskilstuna, Sweden where he has to deal with cost issues in product lines for developing Safety-Critical products at Volvo CE. He has some experience in other Computer Science and Engineering aspects as well and he has been working in several academic and industrial projects during his stay in Sweden. One of them was performed in cooperation with ABB Company in Västerås, Sweden.

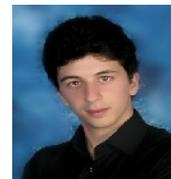

**Miraldi Fifo** was born on July 8th, 1990 in Korçë, Albania. He has graduated the Faculty of Information Technology at the Polytechnic University of Tirana in October 2011. He holds a Bachelor diploma in Computer Engineering from 2011 and he completed the first year of the 2-year master program studies in Computer Engineering (also at the Polytechnic University of Tirana).He has always been interested in Operating Systems, Computer Networks, Software Engineering etc. He presented and defended his Bachelor thesis last year, under the title "Designing and Setting up Servers in Cluster Systems". He was guided by his professor of Computer Networks and Operating Systems, Msc. Igli Tafaj while working with the thesis. He is currently studying for a 2-year master in Software Engineering at Mälardalen University in Sweden.